\documentclass[12pt,preprint]{aastex}

\newcommand{\logg} {\log g}

\newcommand{\hbeta} {H$\beta$}
\newcommand{\Te} {T_{\rm eff}}

\newcommand{\msun} {$M_\odot$}

\newcommand\gta{\lower 0.5ex\hbox{$\buildrel > \over \sim\ $}} 
\newcommand\lta{\lower 0.5ex\hbox{$\buildrel < \over \sim\ $}} 

\shortauthors{Bergeron et al.}
\shorttitle{New ZZ Ceti Stars}
\begin{document}

\title{On the Purity of the ZZ Ceti Instability Strip: Discovery of More 
Pulsating DA White Dwarfs on the Basis of Optical
Spectroscopy\footnote{Based, in part, on observations gathered at the
  European Southern Observatory, La Silla, Chile.}}

\bigskip
\author{P. Bergeron and G. Fontaine}
\affil{D\'epartement de Physique, Universit\'e de Montr\'eal, C.P.~6128, 
Succ.~Centre-Ville, Montr\'eal, Qu\'ebec, Canada, H3C 3J7.}
\email{bergeron@astro.umontreal.ca, fontaine@astro.umontreal.ca}

\author{M. Bill\`eres}
\affil{European Southern Observatory, Santiago Headquarters, Avenida Alonso 
de Cordova 3107, Vitacura, Casilla 19001, Santiago 19, Chile.}
\email{mbilleres@eso.org}

\author{S. Boudreault}
\affil{D\'epartement de Physique, Universit\'e de Montr\'eal, C.P.~6128, 
Succ.~Centre-Ville, Montr\'eal, Qu\'ebec, Canada, H3C 3J7.}
\email{boudreault@astro.umontreal.ca}

\and

\author{E.M. Green}
\affil{Steward Observatory, University of Arizona, Tucson, AZ 85721.}
\email{egreen@as.arizona.edu}

\begin{abstract}

We report the discovery of two new ZZ Ceti pulsators, LP 133$-$144 and
HE 1258$+$0123, selected on the basis of model atmosphere fits to
optical spectroscopic data. The atmospheric parameters for LP
133$-$144, $\Te=11,800\pm 200$~K and $\logg=7.87\pm 0.05$, and for HE
1258$+$0123, $\Te=11,410\pm 200$~K and $\logg=8.04\pm 0.05$, place
them within the empirical boundaries of the ZZ Ceti instability
strip. This brings the number of known ZZ Ceti stars to a total of 36,
a quarter of which have now been discovered using the spectroscopic
approach for estimating their atmospheric parameters. This method has
had a 100\% success rate so far in predicting the variability of
candidate ZZ Ceti stars. We have also analyzed additional spectra of
known nonvariable white dwarfs in the vicinity of the ZZ Ceti
instability strip. Our study further strengthens the idea that ZZ Ceti
stars occupy a pure region in the $\logg-\Te$ plane, a region where no
nonvariable stars are found.  This result supports the thesis that ZZ
Ceti pulsators represent a phase through which {\it all} DA stars must
evolve.

\end{abstract}


\section{Introduction}

Pulsating hydrogen line (DA) white dwarfs --- or ZZ Ceti stars --- are
found in a rather narrow range of effective temperature, between about
$\Te=12,500$~K and 11,100~K according to the detailed study of
\citet[][hereafter B95]{B95}, with the temperatures defining the blue and red edges 
depending on the mass of the white dwarf. Asteroseismological studies of
these stars provide important constraints on their internal structure,
including their chemical layering. If ZZ Ceti stars truly represent an
evolutionary phase through which {\it all} DA white dwarfs must go
through, then the results obtained for the pulsators can be
generalized to the entire class of DA stars as well. In particular,
the asteroseismological determinations of the hydrogen and helium
envelope masses in DA white dwarfs can be used as ``calibration'' of
these quantities in cooling calculations. Hence it is important to
determine the fraction of white dwarfs inside the ZZ Ceti instability
strip that are nonvariable. More than twenty years ago,
\citet{fon82} have argued from a study of multichannel photometric data
that the strip is most likely pure. This question of the purity of the
ZZ Ceti instability strip has been debated in the meantime by various
authors
\citep[e.g.,][]{dolez91,kepler93,kepler95,silvotti97,giovannini98} who
claimed, contrary to \citet{fon82}, that the strip contains both
variable and nonvariable stars. A definite answer to this question had
to wait for a more precise method of measuring the atmospheric
parameters of ZZ Ceti stars and other white dwarfs in the vicinity of
the instability strip.

B95 have developed such a theoretical framework for measuring the
effective temperature and surface gravity of ZZ Ceti stars by
comparing high signal-to-noise ratio (S/N $\gta$ 80) spectroscopic
observations with the predictions of model atmospheres. The method
uses simultaneous fits to the Balmer lines, from H$\beta$ to H9. This
so-called spectroscopic technique has first been applied to hotter
white dwarfs by \citet{BSL} to determine the mass distribution of DA
stars, and it is arguably the most precise method for measuring the
atmospheric parameters of white dwarf stars. Although the approach
used by B95 has been the subject of various criticisms, \citet{fon03}
have reviewed and rebutted all the arguments that have been put
forward against the use of optical spectroscopy. In particular,
\citet{fon03} have demonstrated the usefulness of the spectroscopic
technique to obtain accurate measurements of the atmospheric
parameters of ZZ Ceti stars as well as neighboring stars in the
$\logg-\Te$ diagram, and to predict the variability of ZZ Ceti
candidates. Fontaine et al.~have also shown that the atmospheric
parameters for all 34 known ZZ Ceti stars and 103 nonvariable stars
obtained with the optical spectroscopy approach define a very narrow
region in the $\logg-\Te$ plane in which $no$ nonvariable stars are
found \citep[see Fig.~3 of][]{fon03}, in agreement with our thesis
that ZZ Ceti stars represent a phase through which all DA white dwarfs
must evolve.

With this powerful diagnostic tool in hand, it becomes possible to
identify ZZ Ceti candidates from large white dwarf samples by securing
spectroscopic observations for all objects, and by determining $\Te$
and $\logg$ values from model atmosphere fits to optical
spectroscopy. To obtain reliable and consistent results, however,
three conditions must be met: (1) High-quality spectra must be
gathered, (2) model atmospheres comparable to those of B95 (and based
on the calibration of the mixing-length theory proposed there) must be
used, and (3) attention to details must be provided (see B95 and
references therein). Stars with atmospheric parameters overlapping the
currently known ZZ Ceti stars ought to be variable according to our
past results. There are currently 34 known ZZ Ceti stars, 7 of which
have been discovered using the spectroscopic technique: GD 165
\citep{BM90}, HS 0507+0435B \citep{jordan98}, PG 1541+650
\citep{vauclair00}, GD 244 and KUV 02464+3239 \citep{fon01}, and more
recently MCT 0145$-$2211 and HE 0532$-$5605 \citep{fon03}.  In this
paper we present spectroscopic fits that led to the discovery of two
more pulsators, LP 133$-$144 and HE 1258$+$0123, for a total of 36
ZZ Ceti stars known up to date. We also present an upgraded view of the
ZZ Ceti instability strip in the $\logg-\Te$ diagram.

\section{LP 133$-$144}

\citet{liebert03} have recently obtained high S/N spectroscopy of all DA
white dwarfs identified in the Palomar-Green (PG) survey
\citep{PG86}. One of the goals of that effort is to derive an improved
luminosity function using the spectroscopic approach. Among the 347 DA
stars analyzed in that way, 9 objects were previously known to be ZZ
Ceti variables (GD 99, G117$-$B15A, GD 154, G238$-$53, GD 165, PG
1541$+$651, R808, PG 2303$+$243, and G29$-$38). Among the remaining PG
stars, only one object, LP 133$-$144 (PG 1349$+$552, WD 1349$+$552,
$B_{\rm ph}$ $\simeq$ 16.0), has atmospheric parameters ($\Te=11,800$~K
and $\logg=7.87$) consistent with it being a ZZ Ceti pulsator. Our best
fit using ML2/$\alpha=0.6$ models (see B95) is shown in Figure
\ref{fg:f1}. 

Because Balmer lines in the ZZ Ceti range are, in fact, very sensitive
to both $\Te$ and $\logg$, internal errors obtained from least-squares
fits to high S/N optical spectra are meaningless, as discussed in
detail by \citet{BSL} and B95. The typical good fits that are achieved
only reflect the ability of the model spectra to match the data, and
the error budget is actually dominated by uncertainties of the flux
calibration. To estimate the true external errors, \citet{fon03} have
compared the atmospheric parameters of a subset of ZZ Ceti stars taken
from B95 with those derived from independent spectra obtained by Chris
Moran (2000, private communication), using a completely different
setup and reduction procedure \citep[see also][for a similar
comparison at higher temperatures]{BSL}. As can be seen from Figure 2 of
\citet{fon03}, the standard deviations between both sets of
measurements allow us to estimate the real external errors for stars
in the ZZ Ceti range, $\sigma(\Te)\sim 200$~K and $\sigma(\logg)\sim
0.05$. Since all the stars in this paper are found in the same
temperature range \citep[see Fig.~1 of][]{fon03} and have been
observed with the same setup and signal-to-noise ratio, we adopt
these uncertainties throughout.

To our knowledge, LP 133$-$144 had never been observed before for
photometric variability, most likely because there is no published
color information on this white dwarf (see Fontaine et al. 2001 for a
historical account of the selection criteria used previously for
uncovering ZZ Ceti stars). As part of an ongoing program to identify
new ZZ Ceti pulsators, we observed LP 133$-$144 in integrated (white
light) ``fast'' photometric mode at the Steward Observatory Mount
Bigelow Station 1.6 m telescope during four nights in March-April
2003. The photometric observations were obtained with LAPOUNE, the
portable Montr\'eal three-channel photometer. A total of 17.6 h of
data was gathered. The top two panels of Figure \ref{fg:f2} show our
sky-subtracted, extinction-corrected light curves obtained during the
discovery run. Clearly, LP 133$-$144 is a multiperiodic luminosity
variable, a new ZZ Ceti star.

The top panel of Figure \ref{fg:f3} shows the resulting Fourier
(amplitude) spectrum of the light curve of LP 133$-$144 using all the
17.6 h of data. From this spectrum, we can easily identify four main
frequency components corresponding to periods in the range from 209.2 
to 327.3 s. We note that the relatively short periods and low amplitudes
($\lta 1$ \%) of the luminosity variations detected in LP 133$-$144 are
consistent with its location near the blue edge of the ZZ Ceti
instability strip (see \S~4). 

\section{HE 1258$+$0123}

The two ZZ Ceti stars MCT 0145$-$2211 and HE 0532$-$5605 recently
discovered by \citet{fon03} have published $\Te$ and $\logg$ values
taken from \citet{koester01} as part of the SPY program. These two ZZ
Ceti candidates have been selected by Fontaine et al. because they
happen to fall precisely in the middle of the empirical ZZ Ceti
instability strip determined by B95. Two other SPY objects shown in
Figure 3 of \citet{fon03}, EC 12043$-$1337 (WD 1204$-$136, $V$ =
15.52) and HE 1258$+$0123 (WD 1258$+$013, $V$ = 16.26), fall also
inside the instability strip, but very close to the cool edge of the
strip.  During our March-April 2003 observing run when the variability
of LP 133$-$144 was discovered, we also obtained high speed
photometric data on EC 12043$-$1337, which showed a constant light
curve at the 2$-$3 millimag level. We had also secured similar
observations on HE 1258$+$0123, but we later discovered that we had
observed the wrong star, since the coordinates provided in
\citet{koester01} for this object, $\alpha (2000)=13$:00:59.2 and
$\delta (2000)=+00$:57:12, correspond to the sdO star HE 1258$+$0113
according to \citet{SPY1}, who gives instead for HE 1258$+$0123
$\alpha (2000)=13$:01:10.5 and $\delta (2000)=+01$:07:39.

In June 2003, we secured our own spectroscopic observations for both
EC 12043$-$1337 and HE 1258$+$0123 using the 2.3 m telescope at the
Steward Observatory Kitt Peak Station, equipped with the Boller \&
Chivens spectrograph and a Texas Instrument CCD detector. The spectral
coverage is about $\lambda\lambda$3100--5300, thus covering \hbeta\ up
to H9 at an intermediate resolution of $\sim6$~\AA\ FWHM. Our best fit
for EC 12043$-$1337 using our own model grid is shown in Figure
\ref{fg:f1}. The atmospheric parameters for this object,
$\Te=11,200$~K and $\logg=8.23$, now place it close to, but below, the
empirical red edge of the instability strip, in agreement with our
high speed photometric result. \citet{koester01} obtained
$\Te=11,111$~K and $\logg=8.05$ for the same star.

Our spectroscopic solution for HE 1258$+$0123, $\Te=11,410$~K and
$\logg=8.04$, is shown in Figure
\ref{fg:f1}, which can be compared with the results of \citet{koester01} 
for the same star, $\Te=11,161$~K and $\logg=7.92$.  These revised
parameters push HE 1258$+$0123 even deeper within the empirical
instability strip. On the basis of our previous success at predicting
the variability of ZZ Ceti candidates using the optical spectroscopy
approach, we figured that this star ought to be variable. One week
later, one of us (M.~B.) managed to obtain high speed photometric
observations of HE 1258$+$0123 using the EMMI instrument attached to
the 3.6 m New Technology Telescope (NTT) at the ESO La Silla
Station. The exposure time was adjusted from 20 s to 15 s as the
seeing improved, with corresponding sampling times of 55 s and 49 s,
respectively. Images were bias subtracted, flat fielded, and
magnitudes were calculated using the mag/circ function of the MIDAS
package. The 3.3 hour long photometric light curve is shown in the
lower panel of Figure
\ref{fg:f2}. The results confirm our expectation that HE 1258$+$0123
is indeed a multiperiodic luminosity variable, another ZZ Ceti
star. The Fourier spectrum for this single run is displayed in the
lower panel of Figure
\ref{fg:f3}. The dominant peak at 744.6 s is consistent with this ZZ
Ceti star being closer to the red edge of the instability strip than
LP 133$-$144 (see \S~4). Other important peaks are also present at
439.2, 528.5, and 1091.1 s. The sharp rises and slow declines observed
in the light curve are characteristic of large amplitude ZZ Ceti stars. 

\section{The Empirical ZZ Ceti instability strip}

Of the four SPY objects originally found inside the empirical
instability strip (see Figure 3 of Fontaine et al. 2003), three remain
within the strip according to our own spectroscopic analysis, and they
correspond indeed to new ZZ Ceti stars (MCT 0145$-$2211, HE
0532$-$5605, and HE 1258$+$0123). For its part, the nonvariable white
dwarf EC 12043$-$1337 has a revised effective temperature and a
surface gravity that put it slightly below the empirical red edge of
the strip. Two additional objects lie formally within, but very close
to the {\it blue} edge of the instability strip shown in Figure 3 of
\citet{fon03}. The one at the top is LP 550$-$52 with $\Te=11,550$~K
and $\logg=7.65$, an unresolved degenerate binary with a period of
1.157 days according to
\citet{maxted99}. The atmospheric parameters for this object are thus an
average of the parameters of both components of the system
\citep{liebert91}, and we will no longer consider it in our analysis.
The other object seen in Figure 3 of \citet{fon03} is GD 133 at
$\Te=12,090$~K and $\logg=8.06$. Given the uncertainties, the position
of that object inside the strip, but very close to the blue edge,
certainly remains consistent with the idea of a pure strip. Nevertheless, 
given that the spectrum we used for GD 133 had not been obtained by us,
we reobserved it in June 2003 with our standard setup. Our 
revised atmospheric parameters, $\Te=12,290$~K and $\logg=8.04$, now
place GD 133 above the blue edge of the ZZ Ceti instability strip.

Our updated empirical ZZ Ceti instability strip is shown in Figure
\ref{fg:f4} where the positions of all known variables are indicated,
together with the results for 54 known nonvariable white dwarfs. For
the convenience of the reader, we provide in Table 1 a summary of the
atmospheric parameters for all 36 known ZZ Ceti stars. The table
assembles values taken from B95 (22 stars), Fontaine et al. (2003, 12
stars), and this paper (2 stars). The values of $\Te$ and $\logg$ are
determined from ML2$/\alpha=0.6$ models, while the stellar masses are
derived from the models of \citet{wood95} for carbon-core
compositions, helium layers of $M_{\rm He}=10^{-2} M_{\star}$, and
hydrogen layers of $M_{\rm H}=10^{-4}M_{\star}$. Note that some very
small changes appear with respect to some of the B95 data, the result
of redoing the fits in a completely homogeneous way. The picture of
the empirical instability strip that emerges from our results is that
of a pure strip in which no nonvariable stars are found, a conclusion
that supports our claim that ZZ Ceti stars represent a phase through
which $all$ DA stars must evolve. It has a trapezoidal shape in the
$\logg-\Te$ plane, with the blue edge showing a stronger dependence on
the surface gravity than the red edge does, although this conclusion
rests heavily on the most massive ZZ Ceti star in this diagram, LTT
4816 (WD 1236$-$495).  The newly discovered ZZ Ceti stars reported in
this paper are shown as bold open circles in Figure \ref{fg:f4}, and
their location within the instability strip together with their
dominant periods found from the Fourier spectra (Fig.~\ref{fg:f3}) are
consistent with the period-effective temperature relation observed in
ZZ Ceti pulsators \citep[see, e.g.,][]{winget82}.

In Figure \ref{fg:f4} there appears to be a paucity of stars directly
above the blue edge and below the red edge of the instability
strip. However, our combined sample of variables and nonvariables is
heavily biased against the latter since we have gathered spectra for
only a fraction of them. Progress is underway to secure spectroscopic
observations for all stars outside the boundaries of the instability
strip, and to increase the number of nonvariable white dwarfs in the
vicinity of the strip to help define better the exact location and
shape of the red and blue edges.

The spectroscopic approach, when properly handled, provides 
the most powerful way for discovering new ZZ Ceti stars in a routine 
fashion, and it could be easily applied to large surveys such as the
Sloan Digitized Sky Survey. Interestingly enough, HE 1258$+$0123 is
also part of the Sloan survey (SDSS J 130110.51+010739.9) and its
variability should be rediscovered as part of that ongoing survey.

\acknowledgements We wish to thank the Director and staff of Steward 
Observatory for allowing us to use their facilities. This work was
supported in part by the NSERC Canada and by the Fund NATEQ
(Qu\'ebec). G.F. also acknowledges the contribution of the Canada
Research Chair Program.

\clearpage
\hoffset 0.0truein
\begin{deluxetable}{llcccc}
\tabletypesize{\footnotesize}
\tablecolumns{6}
\tablewidth{0pt}
\tablecaption{Atmospheric Parameters of ZZ Ceti stars}
\tablehead{
\colhead{WD} &
\colhead{Name} &
\colhead{$\Te$(K)} &
\colhead{$\logg$} &
\colhead{$M/$\msun} &
\colhead{$M_V$}}
\startdata
0104$-$464            &BPM 30551           &11260&8.23&0.75&12.16\\
0133$-$116            &R548                &11990&7.97&0.59&11.63\\
0145$-$221            &MCT 0145$-$2211     &11550&8.14&0.69&11.97\\
0246$+$326            &KUV 02464+3239      &11290&8.08&0.65&11.93\\
0341$-$459            &BPM 31594           &11540&8.11&0.67&11.92\\
\\
0416$+$272            &HL Tau 76           &11450&7.89&0.55&11.63\\
0417$+$361            &G38$-$29            &11180&7.91&0.55&11.71\\
0455$+$553            &G191$-$16           &11420&8.05&0.64&11.86\\
0507$+$045            &HS 0507+0435B       &11630&8.17&0.71&11.99\\
0517$+$307            &GD 66               &11980&8.05&0.64&11.75\\
\\
0532$-$560            &HE 0532$-$5605      &11560&8.49&0.92&12.52\\
0836$+$404            &KUV 08368+4026      &11490&8.05&0.64&11.85\\
0858$+$363            &GD 99               &11820&8.08&0.66&11.83\\
0921$+$354            &G117$-$B15A         &11630&7.97&0.59&11.70\\
1137$+$423            &KUV 11370+4222      &11890&8.06&0.64&11.77\\
\\
1159$+$803            &G255$-$2            &11440&8.17&0.71&12.04\\
1236$-$495            &LTT 4816            &11730&8.81&1.10&13.09\\
1258$+$013$^{\rm a}$  &HE 1258+0123        &11410&8.04&0.63&11.84\\
1307$+$354            &GD 154              &11180&8.15&0.70&12.07\\
1349$+$552$^{\rm a}$  &LP 133$-$144        &11800&7.87&0.53&11.53\\
\\
1350$+$656            &G238$-$53           &11890&7.91&0.55&11.56\\
1401$-$147            &EC 14012$-$1446     &11900&8.16&0.70&11.92\\
1422$+$095            &GD 165              &11980&8.06&0.65&11.77\\
1425$-$811            &L19$-$2             &12100&8.21&0.74&11.96\\
1541$+$650            &PG 1541+651         &11600&8.10&0.67&11.90\\
\\
1559$+$369            &R808                &11160&8.04&0.63&11.91\\
1647$+$591            &G226$-$29           &12460&8.28&0.79&12.02\\
1714$-$547            &BPM 24754           &11070&8.03&0.62&11.92\\
1855$+$338            &G207$-$9            &11950&8.35&0.83&12.22\\
1935$+$276            &G185$-$32           &12130&8.05&0.64&11.73\\
\\
\\
\\
1950$+$250            &GD 385              &11710&8.04&0.63&11.78\\
2254$+$126            &GD 244              &11680&8.08&0.65&11.84\\
2303$+$242            &PG 2303+243         &11480&8.09&0.66&11.90\\
2326$+$049            &G29$-$38            &11820&8.14&0.69&11.91\\
2347$+$128            &G30$-$20            &11070&7.95&0.58&11.80\\
\\
2348$-$244            &EC 23487$-$2424     &11520&8.10&0.67&11.91\\
\enddata
\tablenotetext{a}{This paper}
\end{deluxetable}

\clearpage

\figcaption[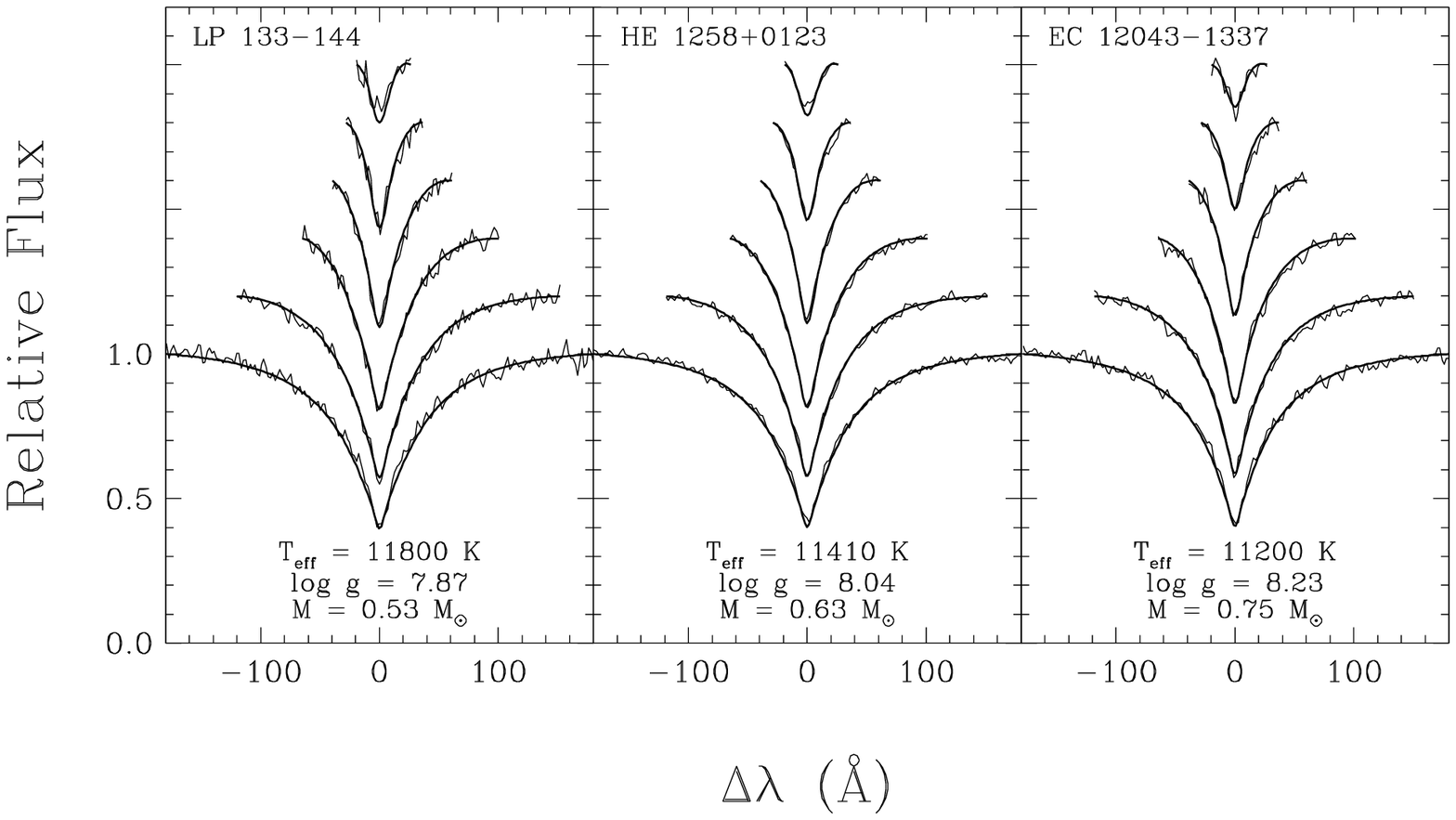] {Model fits to the individual Balmer line profiles
of the new ZZ Ceti stars LP 133$-$144 and HE 1258$+$0123, and of the
nonvariable white dwarf EC 12043$-$1337. The lines range from
H$\beta$ (bottom) to H9 (top), each offset vertically by a factor of
0.2. Values of $\Te$ and $\logg$ have been determined from
ML2$/\alpha=0.6$ models, while the stellar masses have been derived
from the models of \citet{wood95} for carbon-core compositions, helium
layers of $M_{\rm He}=10^{-2} M_{\star}$, and hydrogen layers of
$M_{\rm H}=10^{-4}M_{\star}$. \label{fg:f1}}

\figcaption[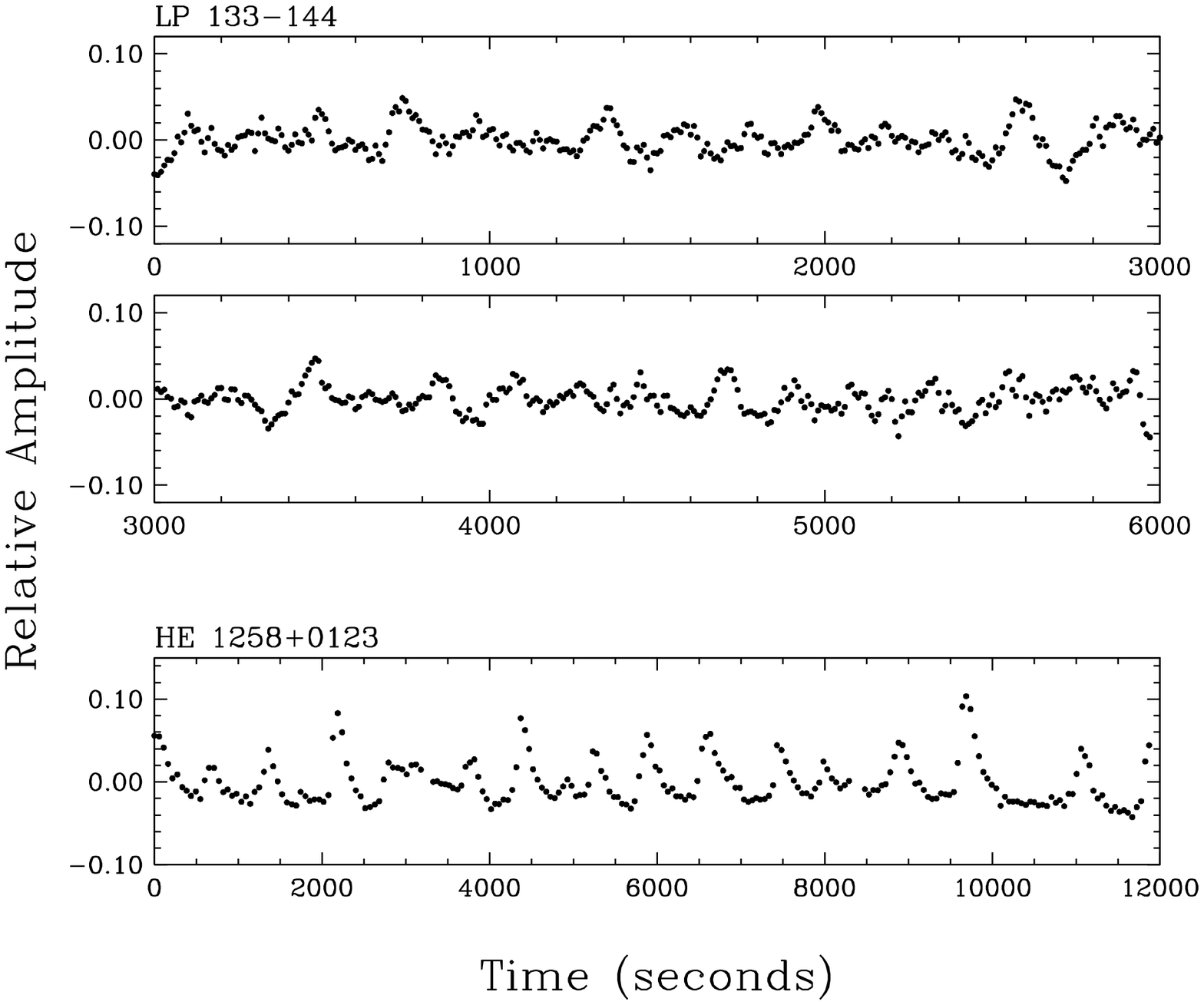] {{\it Top two panels}: Light curve of 
LP 133$-$144, observed in ``white light'' with LAPOUNE attached to the
Mount Bigelow 1.6 m telescope. Each point represents a sampling time of
10 s. {\it Bottom panel}: Light curve of HE 1258$+$0123 gathered using
the EMMI instrument with no filter attached to the NTT. Each plotted
point represents a sampling time of approximately 50 s.  Both light
curves are expressed in terms of residual amplitude relative to the mean
brightness of the star.\label{fg:f2}}

\figcaption[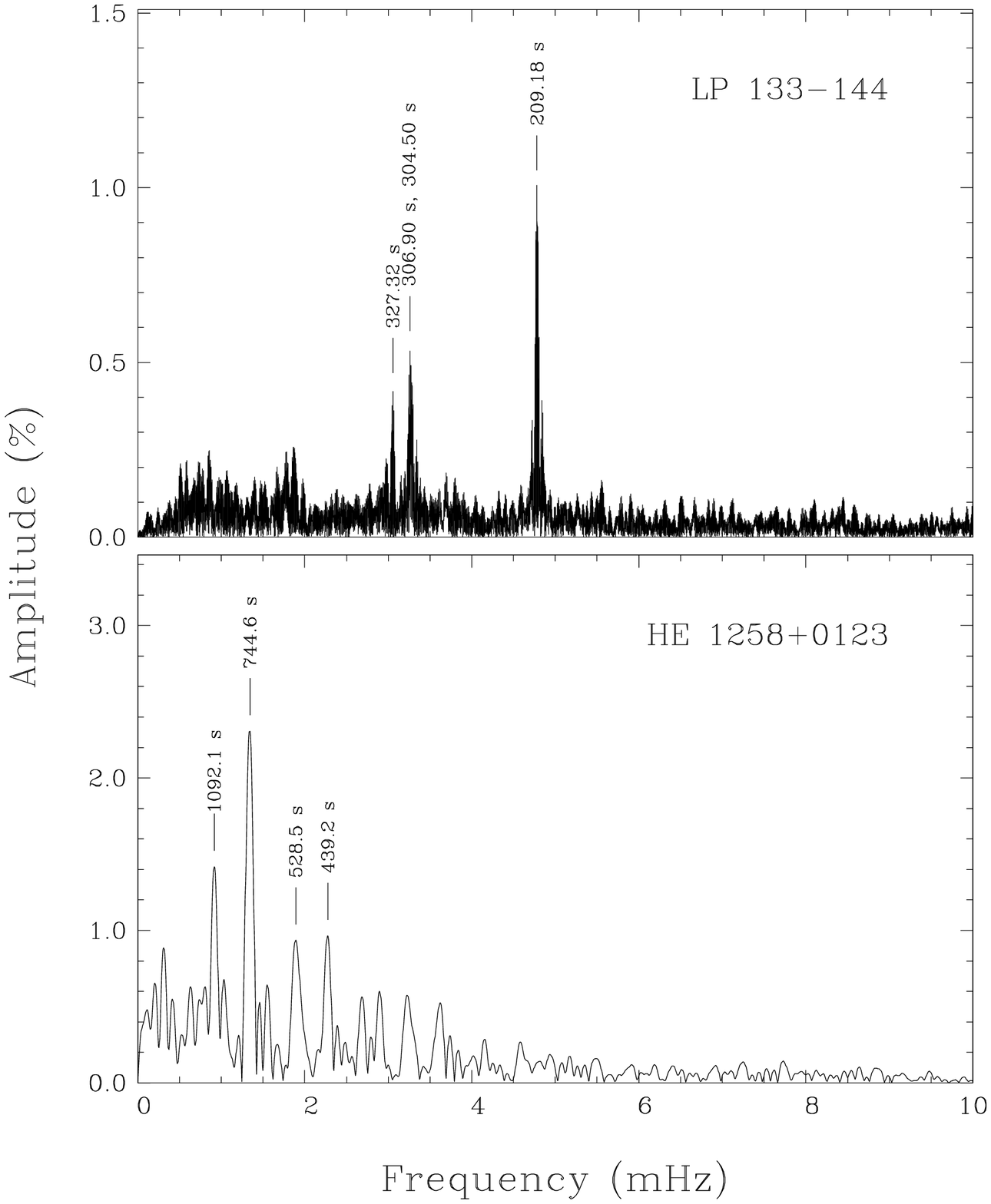] {Fourier (amplitude) spectra of the light curves
of LP 133$-$144 and HE 1258$+$0123 in the 0$-$10 mHz bandpass. The 
spectra in the region from 10 mHz out to the Nyquist frequency are
entirely consistent with noise and are not shown. The amplitude axis is
expressed in terms of the percentage variations about the mean
brightness of the star. \label{fg:f3}}

\figcaption[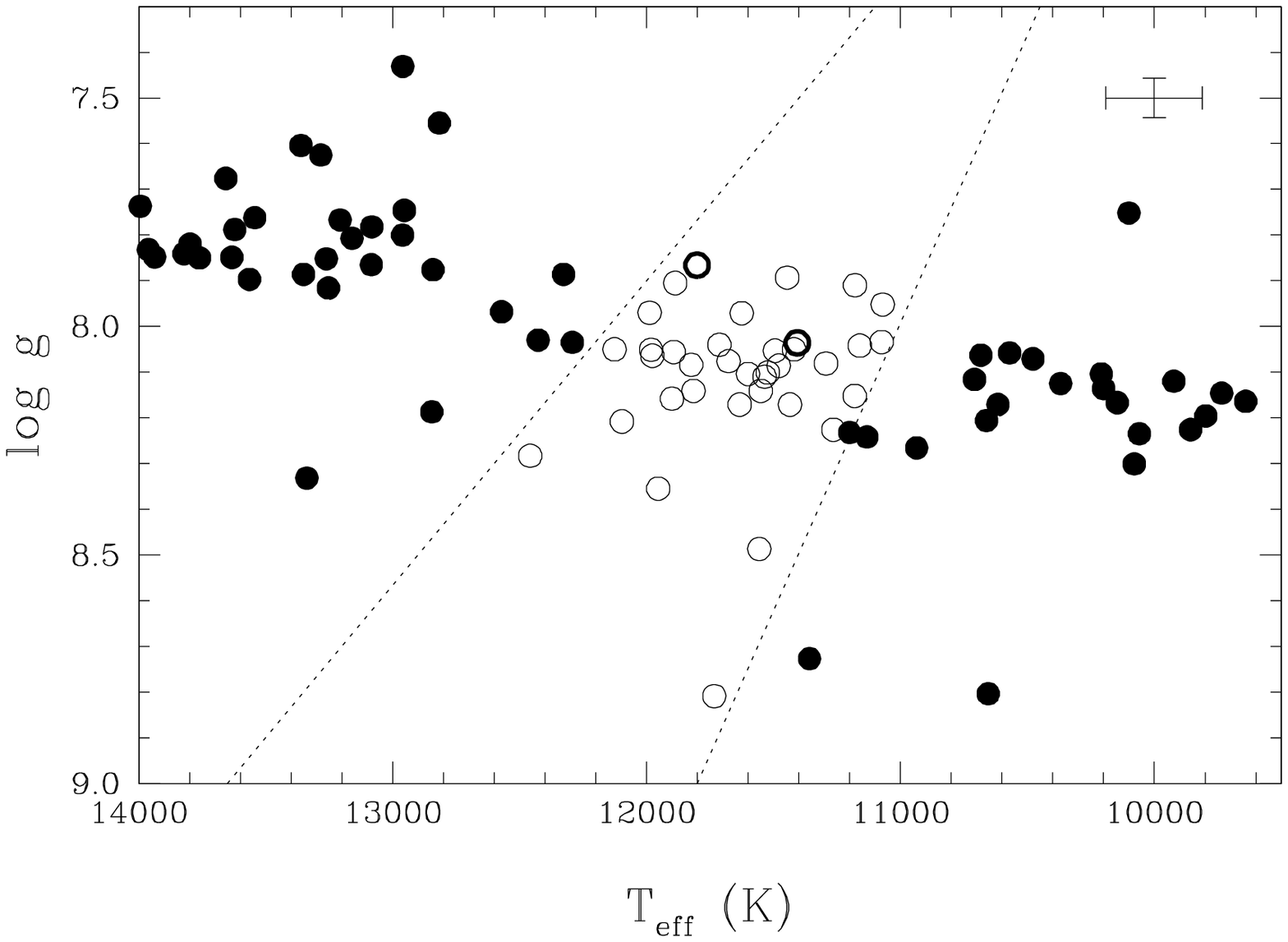] {Surface gravity-effective temperature distribution for
various samples of DA white dwarfs. The open circles represent the 36
known ZZ Ceti stars tabulated in Table 1; the bold open circles
correspond to the newly identified ZZ Ceti stars LP 133$-$144 ({\it
left}) and HE 1258$+$0123 ({\it right}). Filled circles are DA stars
that are known to be nonvariable and whose atmospheric parameters have
been determined by us on the basis of the same homogeneous approach as
the ZZ Ceti stars. The error bars correspond to the uncertainties of
the spectroscopic method in the region where ZZ Ceti stars are
located, $\sigma(\Te)\sim 200$~K and $\sigma (\logg)\sim 0.05$,
as estimated by \citet{fon03}. The dashed lines represent the
empirical blue and red edges of the instability strip. \label{fg:f4}}

\clearpage
\begin{figure}[p]
\plotone{f1.eps}
\begin{flushright}
Figure \ref{fg:f1}
\end{flushright}
\end{figure}

\clearpage
\begin{figure}[p]
\plotone{f2.eps}
\begin{flushright}
Figure \ref{fg:f2}
\end{flushright}
\end{figure}

\clearpage
\begin{figure}[p]
\plotone{f3.eps}
\begin{flushright}
Figure \ref{fg:f3}
\end{flushright}
\end{figure}

\clearpage
\begin{figure}[p]
\plotone{f4.eps}
\begin{flushright}
Figure \ref{fg:f4}
\end{flushright}
\end{figure}

\end{document}